\documentclass[12pt]{iopart}
\expandafter\let\csname equation*\endcsname\relax
\expandafter\let\csname endequation*\endcsname\relax
\usepackage{iopams}  
\usepackage{amsmath,amssymb}
\usepackage[countmax]{subfloat}
\usepackage{graphicx}
\usepackage[usenames]{color}
\usepackage{hyperref}

\newcommand{\erefo}[1]{Eq.~(\ref{#1})}
\newcommand{\re}[1]{(\ref{#1})}
\newcommand{\esref}[1]{Eqs.~(\ref{#1})}

\newcommand{\beg}{\begin{equation}}
\newcommand{\en}{\end{equation}}

\newcommand{\eps}{\epsilon}

\newcommand{\bsub}{\begin{subequations}}
\newcommand{\esub}{\end{subequations}}

\begin{document}

\title{Integrals of motion for one-dimensional Anderson localized systems}

\author{ Ranjan Modak$^1$ , Subroto Mukerjee$^{1,2}$, Emil A. Yuzbashyan$^{3}$ and B. Sriram Shastry$^4$,  }

\address{$^1$ Department of Physics, Indian Institute of Science, Bangalore 560 012, India\\
 $^2$ Centre for Quantum Information and Quantum Computing, Indian Institute of Science,
Bangalore 560 012, India\\
$^3$Center for Materials Theory, Rutgers University, Piscataway, NJ 08854, USA\\
 $^4$Physics Department, University of California, Santa Cruz, CA 95064, USA}

\begin{abstract}
 Anderson localization is known to be  inevitable  in one dimension  for generic disordered models. Since localization leads to Poissonian energy level statistics,  we ask if localized systems possess ``additional''  integrals of motion as well, so as to enhance the analogy with quantum  integrable systems. We answer this in the affirmative in the present work.  We construct a set of nontrivial integrals of motion for Anderson localized models,  in terms of the original creation and annihilation operators. These are found   as a power series in  the  hopping  parameter. 
   The recently found Type-1 Hamiltonians, which are known to be quantum integrable in a precise sense, motivate our construction.  We note that these models  can be viewed as  disordered electron  models with {\em infinite-range} hopping, where  a similar  series  truncates at the linear order. We show that  despite the  infinite range hopping, all states but one  are localized.    We also study the conservation laws for the disorder free Aubry-Andre model, where the states are either localized or extended, depending on the strength of a coupling constant. We formulate  a specific  procedure  for averaging over  disorder, in order to examine the convergence of the power series. Using this procedure  in the Aubry-Andre  model,  we show that  integrals  of motion given by our   construction are well-defined  in localized phase, but  not so  in the extended phase.   Finally, we also obtain the integrals of motion for a model with interactions to lowest order in the interaction.  
\end{abstract}


\maketitle
\section{Introduction}
The simplest theoretical model to study localization for non-interacting particles in the presence of disorder was proposed by Anderson~\cite{anderson.1958}.  A single particle localized state has a wavefunction that decays exponentially about some point in space over a characteristic localization length. In three dimensions, localized states exist below a certain energy (the mobility edge) for a given strength of disorder. A disordered electronic system is thus localized if its Fermi energy lies below the mobility edge. In one and two dimensions, an infinitesimal amount of disorder is sufficient to localize all single particle states and thus a disordered non-interacting electronic system is always localized~\cite{tvr.1985,tvr.1979}. 

Recent developments in the area of eigenstate thermalization \cite{deutsch,srednicki,rigol.2007}  relate closely to the above well established notions of Anderson localization.
In this context, it is believed that an isolated localized eigenstate does not thermalize,  in the sense that no subsystem of it can be brought into thermal equilibrium by exchanging heat with the rest of the system. An analogous statement can be made about  information, as defined through an appropriate partial trace of the density matrix.   A related feature of such a system is the lack of level repulsion in its energy level spectrum. This can be thought of as arising from the presence of almost degenerate states localized so far apart that they are unable to hybridize to lift the degeneracy. 

The effect of interactions on such systems is very interesting. Interactions among the elementary degrees of freedom generically tend to drive the system towards thermalization and 
delocalization~\cite{basko.2006,huse.2015}. This tendency competes with the the one that causes localization in the presence of disorder. Understanding the resultant phenomenon of many body localization, that is observed for sufficiently strong disorder, is currently a very active area of research~\cite{pal.2010,vosk.2013,vadim.2007,bardarson.2012,agarwal.2014}.

 Another class of systems that fail to thermalize are integrable  ones. These often contain a variable parameter (such as interaction or external field strength, which we denote here as $y$)  and possess a set of similarly dynamical 
 (i.e.  depending on the parameter) integrals of motion. Standard examples of such systems are the 
 one-dimensional Hubbard and
 XXZ models.  In  these examples, the integrals of motion $I_k$ 
 are polynomial in $y$ with the order of the polynomial~\cite{sriram.1986,shastry.1986,luscher,grosse,grabowski.1995,gould,tang1990,Fuchssteiner} increasing with $k$. An arbitrary linear 
 superposition of all integrals $Q=\sum_k a_k I_k$ -- also an integral in its own right -- is  an 
 infinite power series in $y$.  
 Gaudin magnets\cite{relano,sklyanin1} on the other hand provide  examples of  integrable  models where all conserved charges\footnote{We use the term conserved  charges interchangeably with  conservation laws or integrals of motion.} are linear in $y$. It should be emphasized that there is no generally accepted precise notion of   integrability   in quantum mechanics~\cite{caux,emil.2013} in contrast to classical mechanics where it is unambigous. However, we do not dwell on this issue in present work\footnote{By integrable we will generally mean quantum   many-body models colloquially recognized as such, see  examples in this paragraph. The only exception are type-1 Hamiltonians  that stem from a recently proposed well-defined notion of quantum integrability~\cite{shastry.2005,owusu,shastry2011,owusu2011,emil.2013,owusu2013functionally}.}. The only aspect  important for us here is the  existence of  parameter-dependent conservation laws.

Conserved charges   greatly constrain the dynamics of integrable systems. As a result, when started off from an arbitrary initial state in isolation, these systems do not evolve in a way that causes thermalization in the sense of the above paragraph~\cite{rigol.2007,rigol.2009}. Additionally the usual space time symmetries  result in degeneracies in the energy level spectrum,  and hence a lack of level repulsion~\cite{berry.1977}. The addition of perturbations destroys such conservation laws and restores level repulsion, although the strength of the perturbations has a non-trivial finite-size dependence~\cite{ranjan.2014,subroto.2014,rabson.2004}.   

 In this context, it is natural to ask  in what ways are localized systems similar to integrable ones. In particular we may ask if (parameter dependent) conservation laws, similar to those in integrable systems exist for localized systems. It has been argued in the context of many-body localization that they do, and results related to the growth of entanglement in these systems are predicated on their existence~\cite{huse2014phenomenology,serbyn2013local,dima.2015}. However, obtaining the structure of the conserved charges directly in terms of microscopic parameters remains a challenge and effective renormalization procedures need to be employed instead~\cite{vosk.2014,vosk.2013}. The situation is less complicated in the absence of interactions since the Hamiltonian is that of a single particle. Nevertheless, obtaining the charges systematically and analytically in terms of the microscopic parameters of the Hamiltonian is non-trivial. In this paper  we outline the procedure to do so. We also elucidate the connection between localization and conserved charges.

In this work we study a general  one dimensional model with on-site disorder that can interpolate between models with long-range hopping and the 
more standard Anderson-type one. The starting point is a   Type-1 Hamiltonian reviewed in Refs.~\cite{shastry.2005,owusu,shastry2011,owusu2011,emil.2013,owusu2013functionally}.  This was introduced as the most simple  model of quantum integrability in finite dimensional spaces. This model has infinite ranged hopping, and as such has no inbuilt metric or length scale. We first  show  by calculating its   Participation ratio 
(PR)~\cite{weaire.1977,johri.2014} the  perhaps surprising  result  that all states except  one  are localized.
  This is done as follows:  an eigenstate $|\psi \rangle$ of the Hamiltonian is expanded in a basis of position eigenstates on the lattice as 
$| \psi \rangle = \sum_k c_k |k\rangle$, 
where $k$ labels the position eigenstates and $c_k$ are the coefficients in the expansion. 
The PR for this state is then defined as
\begin{equation}
\mathrm{PR}_{\psi }=\frac{\left(\sum_{k}|c_k|^{2}\right)^2}{\sum_{k}|c_k|^{4}}.
\label{pr}
\end{equation}
It is usually understood that $\mathrm{PR}_\psi \sim O(1)$ indicates localization while  $\mathrm{PR}_\psi\sim O(N)$ -- delocalization, where $N$ is    the number of sites. While this definition is valid for a fixed wave function, we may also define the PR at a given energy, as  later in the paper, where an averaging over disorder realizations is carried out, at a fixed energy.

 The Type-1 model has  a known set of conservation laws, which inspire the construction of a generic  Anderson-type model having only nearest-neighbor hopping. In 1d it is well known that for this model,
 all  single particle eigenstates are localized for any strength of the disorder. The conserved charges of this model are then constructed by analogy with the Type-1 Hamiltonian.  These charges are  expressed as a power series in the hopping, whose coefficients we  determine by means of an algorithm. We also show that the series, upon disorder averaging over a  ``non-resonant''  ensemble- defined below,  is convergent. This  provides numerical  evidence that the ensemble chosen and the procedure of averaging the coefficients in the  conserved charges over the ensemble  is   meaningful. 
 
 We then turn our focus to a model which contains both localized and delocalized phases (i.e. phases in which {\em all} single particles states are either localized or delocalized). This is the Aubry-Andre model~\cite{aubry.1980}, in which the random potential is replaced by a quasi-periodic one. This allows us to test our criterion for the convergence of the power series and clearly elucidate the connection of the conserved charges to localization. Thus, the  convergence (divergence) of the power series representation of conserved charges can indeed be identified with the presence (or absence) of localization and the localization-delocalization transition can be located using the charges. Finally, we investigate the effect of interactions and argue that a power series in the interaction becomes intractable and thus obtain the the conserved charges only to first order in it.
 
 We emphasize that the main feature of our construction is that the conservation laws do not depend explicitly on the wavefunctions of the single particle energy eigenstates. In fact, the recursion relations we obtain for the coefficients of the expansions of the conserved charges are the same for {\em all} generic one dimensional models. Our approach is thus completely model independent requiring no knowledge of exact solutions or properties of energy eigenfunctions. 

 Another important aspect of the construction of conservation laws we emphasize here, which has not been discussed before is `gauge freedom'  of a certain kind, defined more precisely later. We show that a judicious choice of gauge can bring out important features of the conserved charges, such as the truncation of their series representation at finite order. These features can be obscured in gauges that arise in constructions of these charges from direct applications of standard methods such as the Rayleigh-Schr\"{o}dinger series or the locator expansion.

\section{Lattice Models}
We consider a general Hamiltonian  of non-interacting particles hopping on a one dimensional lattice with an on-site potential
 \begin{equation}
H=H(y)= \sum_{i}{\epsilon_{i}n_{i}}-y\sum_{ij}{t_{ij}c_{i}^{\dag}c_{j}},
\label{Eq:genham}
\end{equation}
where $c_{i}^{\dag}$ and $c_{i}$ are fermionic creation and annihilation operators, $n_{i}=c_i^\dag c_i$ is the number operator,
$\epsilon_i$ is the on-site disordered potential, and $t_{ij}$ is the hopping between sites $i$ and $j$. The parameter $y$ is a real number introduced for convenience, which; it allows us to perform an expansion of the conserved charges in its powers.

 Our general strategy to construct construct conserved charges for this models will be to first consider the `unperturbed' Hamiltonian which only has the on-site potential. The conserved charges for this Hamiltonian are simply the operators $n_i$, which are independent and commute  with each other and the Hamiltonian. It can also be readily seen that the eigenstates of this Hamiltonian are completely localized on the individual sites. Thus the zeroth order Hamiltonian trivially describes a localized system with conserved charges. We now show that upon introducing the hopping, new conserved charges $Q_i$ appear, which can still be labeled by the site indices $i$ while the system remains localized. To do this, we consider different types of hopping parameters $t_{ij}$.

\section{Type-1 \protect Hamiltonians}
 We now summarize  a known set of conserved charges $Q_j$. We rework the  construction in Refs.~\cite{owusu,shastry2011,emil.2013}, in a fashion  that  suggests a natural  generalization for short ranged models. These charges    are linear in the hopping (or the parameter $y$), and   commute exactly with the Hamiltonian of the Type-1 family. The Type-1 Hamiltonian is   obtained from Eq.~(\ref{Eq:genham}) by specializing to infinite ranged  hopping $t_{ij}= \gamma_i \gamma_j$, with arbitrary parameters $\gamma_j$.  Specializing to $j=0$ we write down the charge $Q_0$
\begin{eqnarray}
Q_{0}&=&n_{0}-y\sum_{k \neq 0}{\frac{1}{\epsilon_{0}-\epsilon_{k}}[t_{0k}(c_{0}^{\dag}c_{k}+c_{k}^{\dag}c_{0})\underbrace{-\alpha_{k}^{0}n_{0}
-\beta_{k}^{0}n_{k}}]} ,\nonumber \\
\label{Eq:genqtypeI}
\end{eqnarray}
where $\alpha_k^0$ and $\beta_k^0$ are yet to be determined.The commutator of $Q_{0}$ and $H$ vanishes to linear order in $y$ by construction.The surviving term is of $O(y^2)$ and is  given by
\begin{eqnarray}
[Q_{0},H]=y^{2}\sum_{jk}\frac{1}{\epsilon_{0}-\epsilon_{k}}&&[A(0,j,k)(c_{0}^{\dag}c_{j}-c_{j}^{\dag}c_{0})
 \nonumber \\
&+&B(0,j,k) (c_{k}^{\dag}c_{j}-c_{j}^{\dag}c_{k})]  =0,
\label{Eq:comtypeI}
\end{eqnarray}
where
\begin{eqnarray}
 A(0,j,k)=t_{0k}t_{kj}-\alpha_{k}^{0}t_{0j} \nonumber \\
 B(0,j,k)=t_{0j}t_{0k}-\beta_{k}^{0}t_{kj}. \nonumber \\ \label{abvanish}
\end{eqnarray}
 A few words on the form of Eq.~(\ref{Eq:genqtypeI}) are appropriate here.  The last two terms $-\alpha_{k}^{0}n_{k}
-\beta_{k}^{0}n_{0}$ commute with $H({y=0})$ trivially, since they are expressed in terms of the number operators. These actually represent a particularly convenient ``gauge choice'', their presence enables the second order term  $O(y^2)$ to vanish,  and thus the commutator series to truncate exactly for the Type-1 matrices.
The requirement that $[Q_0,H]=0$ is satisfied by the following form of $t_{ij}$.
\begin{eqnarray}
t_{ij}=\gamma_{i}\gamma_{j} \nonumber \\
\alpha_{j}^{0}=\gamma_{j}^{2} \nonumber \\
\beta_{j}^{0}=\gamma_{0}^{2}, 
\label{Eq:typeIdef}
\end{eqnarray}
 this gives $A(0,j,k)=B(0,j,k)=0$. It is straightforward to extend this definition to arbitrary $Q_j$, and further to show that
$[Q_i,Q_j]=0~\forall i,j$, so the operators $Q_i$ are indeed the conserved charges of the Hamiltonian $H$\cite{shastry.2005,owusu}. The Hamiltonians described by $t_{ij}$ of the form given in~\erefo{Eq:typeIdef} are called Type 1~\cite{owusu,emil.2013}, and can also be interpreted  geometrically as representing  a `d-simplex'~\cite{ossipov}. 

\subsection{PR for Type-1 Hamiltonians}

\begin{figure}
\begin{center}
\setlength{\unitlength}{8.2cm}
\begin{picture}(1, 0.718)(0,0)
   \put(0,0){\resizebox{1\unitlength}{!}{\includegraphics{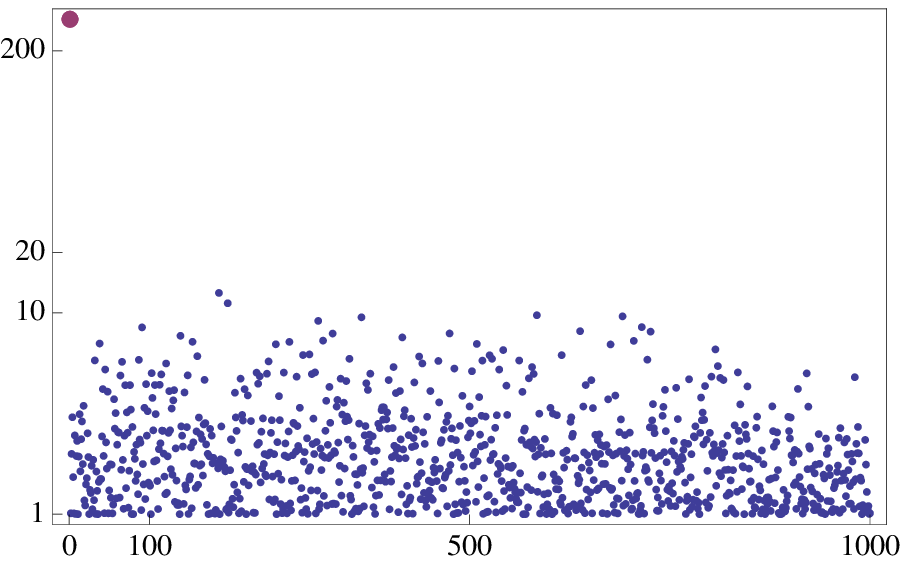}}}
\put(0.11,0.37){\resizebox{0.325\unitlength}{!}{\includegraphics{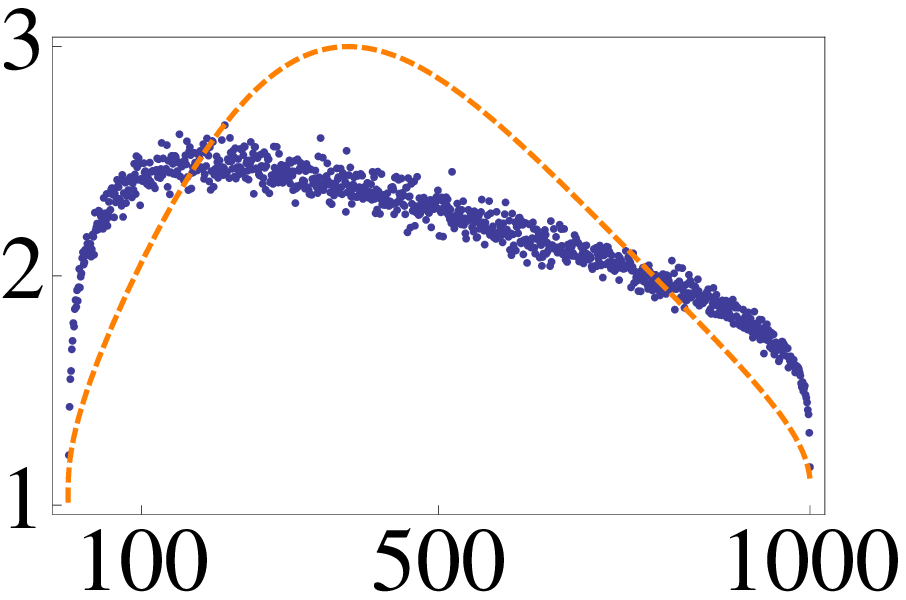}}}
\put(0.62,0.37){\resizebox{0.325\unitlength}{!}{\includegraphics{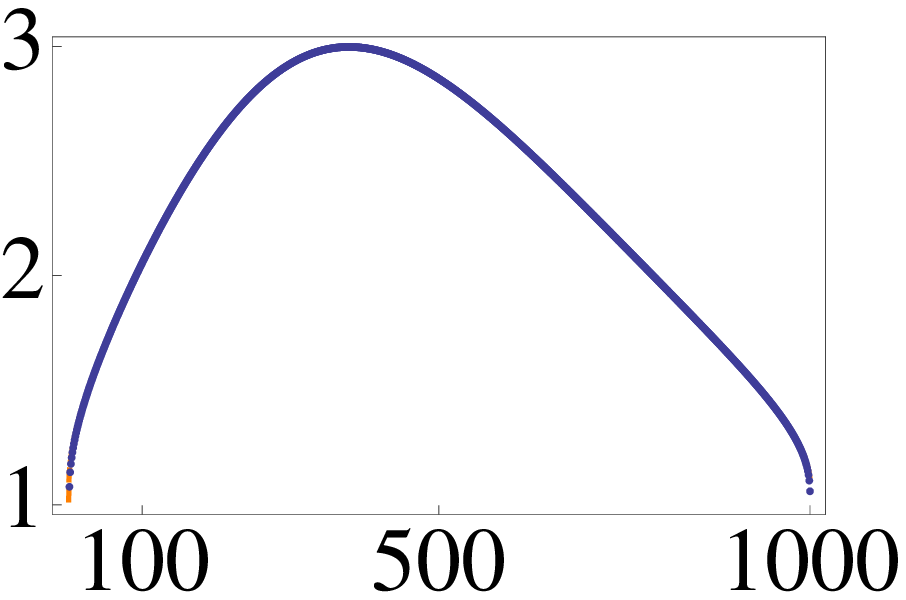}}}
    \put(0.45,-0.03){\makebox(0,0)[b]{\bf Eigenstate}}
   \put(-0.01,0.44){\rotatebox{90}{\makebox(0,0)[t]{\bf PR}}}
\end{picture}
\end{center}
\caption{ PR of eigenstates of a Type-1 Hamiltonian   for  $ y=.004$, $N=10^3$    in ascending order according to the energy.   Each $\epsilon_i$ and $\gamma_i$ is an   independent
  random number  uniformly distributed in an interval $(-1,1)$. Larger circle near the left top corner indicates the ground state, which is extended. Left inset is the same as above, but averaged over $10^3$   realizations of disorder and compared to \erefo{prex} for the same $y,N,w$. The right inset shows the PR (except the ground state) for $N=10^3$ equally spaced $\eps_i$, $\gamma_i=1$ and $ y=.004$ similarly compared to \erefo{prex} (the two curves are indistinguishable). }
\label{Fig:PR_type1}
\end{figure}

All single particle states of Type-1 Hamiltonians \re{Eq:typeIdef}, except  possibly the ground state for $y>0$ or the highest energy
state for $y<0$ are localized, see e.g.  
  Fig.~\ref{Fig:PR_type1}. 
 
 This can be understood in more detail from the exact solution for the spectrum of these models\cite{owusu}. Exact un-normalized single particle eigenstates of the  Hamiltonian \re{Eq:typeIdef}
 are
 \beg
 |E\rangle=\sum_{i=0}^{N-1} \frac{\gamma_i c_i^\dag}{E-\eps_i} |0\rangle,
 \label{state}
 \en
and the corresponding eigenvalues $E$ (energies) are
 solutions of the  equation
 \beg
\sum_{i=0}^{N-1} \frac{\gamma_i^2}{E-\eps_i}=-\frac{1}{y}.
\label{ex}
\en
Suppose $\eps_i$ are ordered in the ascending order. By plotting the left hand side of  \erefo{ex} as a function of $E$, one can verify that it has $N-1$ real roots $E_1, E_2,\dots E_{N-1}$ located between consecutive
$\eps_i$, i.e.  $\eps_{i-1} <E_i<\eps_i$. The remaining root $E_0$ is also real and is below $\eps_0$ (ground state) for $y>0$ and above $\eps_{N-1}$ for $y<0$ (highest excited state).

 \esref{state} and \re{ex} also provide an exact solution for one fermion (Cooper) pair and one spin flip   sectors of the BCS and Gaudin models, respectively,
\beg
\begin{split}       
H_\mathrm{BCS} = \sum_{i, \sigma=\uparrow, \downarrow}\!\!\!  \epsilon_i c^\dagger_{i\sigma} c_{i\sigma} -   y\sum_{ij}
 c^\dagger_{i\downarrow} c^\dagger_{i\uparrow}  c_{j \uparrow} c_{j\downarrow},\\
 H_i(x)= s_i^z-y\sum_{j\ne i}\frac{\vec s_i\cdot \vec s_j}{\eps_i-\eps_j},\\
 \end{split}
 \label{bcs}
\en
where $c_{i\sigma}$ are spin-full fermions and $\vec s_i$ are quantum spins of arbitrary magnitudes $s_i$, see Ref.~\cite{owusu} for details. For the BCS [Gaudin] model one needs to replace
$c_i^\dag \to c^\dagger_{i\downarrow} c^\dagger_{i\uparrow}$ [$s_i^+$] in \erefo{state}, set  $\gamma_i=1$ [$\sqrt{2s_i}$] and the corresponding eigenvalue is equal to $2E$ [$2s_i (E-\eps_i)^{-1}$] rather than $E$. Our results for the PR of Type-1 Hamiltonians therefore also apply to these sectors of these models.

The PR defined through \erefo{pr}  reads
\beg
\mathrm{PR}_E=\frac{\left[ \displaystyle \sum_i \frac{\gamma_i^2}{(E-\eps_i)^2}\right]^2 } {\displaystyle \sum_i \frac{\gamma_i^4}{(E-\eps_i)^4} }.
\label{pr1}
\en
For concreteness we take $y>0$. Then, the ground state is $E_0<\eps_0$. We  assume that most $\gamma_i$ are of the same order of magnitude and consequently the vector with components $\gamma_i$ is delocalized. Further, we take $\eps_i$ to lie in a fixed interval that does not scale with $N$, e.g. from $-w$ to $w$. 

For excited states $E_k$ is between $\eps_{k-1}$ and $\eps_k$. The summations in the numerator and denominator of \erefo{pr1} both  come from $\eps_i$  in a small vicinity of $\eps_k$ for large $N$ and converge as $\sum_n n^{-2}$ and $\sum_n n^{-4}$, respectively, where $n=|i-k|$. The numerator and the denominator scale as $[\gamma_k^2/\delta^2]^2$ and $\gamma_i^4/\delta^4$, where $\delta\propto 1/N$ is the mean level spacing between $\eps_i$ in the vicinity of $\eps_k$. Therefore, $\mathrm{PR}_{E_k}$ is of order 1 (much smaller than $N$)   meaning  excited states are always localized. Fig.~\ref{Fig:PR_type1} shows PR for $N=10^3$ uncorrelated random $\eps_i$   uniformly drawn from an interval $(-1, 1)$ and the same distribution of $\gamma_i$. 

Consistent with our numerical results, we estimate the largest PR for excited states to scale as $\ln N$, i.e.
\beg
\mathrm{PR}_{E_k}^{\max}\approx \alpha\ln N,
\label{log}
\en
for large $N$, where $\alpha$ depends on $N$ much weaker than $\ln N$. Such values of PR come from clustering in $\eps_i$. Indeed, suppose spacings $\delta_i=\eps_{i+1}-\eps_i$ between $m$ of $\eps_i$ for $i$ from $k$ to $k+m$ are all much smaller than $\delta_{k-1}$ and, moreover, $\eps_{k+m}-\eps_k\le \delta_{k-1}$. It follows from \erefo{pr1} that $\mathrm{PR}_{E_k}\approx
\mathrm{PR}_{E_{k+m+1}}\approx m$ because the above $\eps_i$ contribute most to these PRs. Normalized spacings $s_i=\delta_i/\delta$ are distributed according to the Poisson distribution $P(s) ds = e^{-s} ds$. The probability of having $m$ spacings between 0 and $s_0\ll 1$ is then roughly $s_0^m$. We need $m s_0\le 1$ 
and also $N s_0^m =1$ so that at least one such clustering occurs\footnote{ More precisely, the probability that $m$ of $\eps_i$ occur in an interval of length  $\delta$ for Poisson distribution
is $e^{-1}/m!$, which however still leads to the same estimate \re{log} } . This implies $m\approx \ln N/\ln(\ln N)$ and \erefo{log} follows. 
Numerically we find that typical values of $\alpha\approx 1- 3$   and averaged over disorder $\bar\alpha\approx 1.7$, at least for $N=2^4 - 2^{12}$.
Note that according to this argument such large values of PR typically come in pairs  spaced by $m+1$, roughly equal 
to the value of the PR itself. We also stress that, in contrast to the largest PR, a typical (and average) 
  PR is something between one and three for any $N$ (does not scale) as can be seen from  Fig.~\ref{Fig:PR_type1}.

It is interesting to compare this $\ln N$ behavior to the flat band localization 
studied earlier~\cite{nishino2007flat,chalker2010anderson}. The latter  leads 
  to a (weakly) divergent  PR  in  the localized regime, a phenomenon that is viewed as corresponding to critical (power law type) 
  localization.  
 The Type-1 Hamiltonian kinetic energy may also be viewed as a ``flat band'' model, with a flat 
  dispersion for all  except  one state.  Indeed, for $t_{ij}=\gamma_i\gamma_j$ all but one eigenvalues of 
  the second term in \erefo {Eq:genham} are  zero.  The non-zero eigenvalue (ground state for $y>0$) 
  corresponds to the eigenstate $\gamma_i c_i^\dag |0\rangle$. 

Let us consider limits $y\to 0$ and $y\to\infty$ separately. When $y\to0$ all states are localized as expected. Indeed, \erefo{ex} implies $E_k \to \eps_k$, summations in \erefo{pr1} are dominated by the $i=k$ term and we obtain $\mathrm{PR}_{E_k}=1$ for all $k$. When $y\to\infty$ excited states are localized as before because $E_k$ for $k\ge 1$ remains trapped in the interval $(\eps_{k-1}, \eps_k)$. The ground state energy on the other hand diverges -- \erefo{ex} implies $E_0\to -y\sum_i\gamma_i^2$. Then, $\eps_i$ are negligible as compared to $E_0$ in \erefo{pr1} and 
\beg
\mathrm{PR}_{E_0}=\frac{\left[  \sum_i  \gamma_i^2 \right]^2 } {  \sum_i  \gamma_i^4  },
\label{pr0}
\en
which is of order $N$ according to our choice of $\gamma_i$. The ground state is therefore delocalized for $y\to \infty$. It undergoes a localization-delocalization crossover at a certain $y_c$,  which we estimate below in this section.

It is possible to evaluate the PR   analytically to leading order in $1/N$ for distributions of $\eps_i$ and $\gamma_i$  with negligible short range fluctuations (such that
the spacing $\delta_i=\eps_{i+1}-\eps_i$  changes slowly with $i$ -- $|\delta_{i+1}-\delta_i|/\delta_i$ is of order $1/N$ for all $i$ -- and similarly for $\gamma_i$). For simplicity,  let us take constant $\gamma_i$, which we can set to one with no loss of generality, and equally spaced $\eps_i$, i.e. $\delta_i= \delta=2w/N$. 

For excited states, we write $E_k=\eps_{k}-\alpha_k \delta$, where $0<\alpha_k <1$, and solve \erefo{ex} for $\alpha_k$ to the leading order in $1/N$ as described in Appendix B of\cite{yuzbashyan2005}. This yields
\beg
\cot \pi\alpha_k  =\frac{\delta}{\pi y}+\frac{1}{\pi}\ln\frac{\eps_k+w}{w-\eps_k}\equiv f(\eps_k).
\label{alpha}
\en
We note that $\lambda=y/\delta$ is the proper dimensionless coupling constant in the sense that it must stay finite in the $N\to\infty$ limit. This is because the second summation in 
\erefo{Eq:genham} scales as $N^2$  for $t_{ij}=\gamma_i\gamma_j$ and our choice of $\gamma_i$. Therefore, we need $y\propto\delta\propto 1/N$ so that both terms  in 
\erefo{Eq:genham} are extensive in the thermodynamic limit. For the BCS Hamiltonian in \erefo{bcs}, so defined $\lambda$ is the dimensionless superconducting coupling\cite{delft}.

\erefo{pr1} becomes to leading order in $1/N$
\beg
 \mathrm{PR}_{E_k}=\frac{ \left[ \sum_{n=0}^\infty\left(\frac{1}{(n+\alpha_k)^2}+
 \frac{1}{(n+1-\alpha_k)^2}\right)  \right]^2 } { \sum_{n=0}^\infty\left(\frac{1}{(n+\alpha_k)^4}+
 \frac{1}{(n+1-\alpha_k)^4}\right)},
\label{pr}
\en
which evaluates to
\beg
 \mathrm{PR}_{E_k}= \frac{3}{1+2\cos^2\pi\alpha_k}=\frac{3+3f^2(\eps_k)}{1+3f^2(\eps_k)}.
 \label{prex}
 \en
This answer is in good agreement with numerics already for $N=20$, see also Fig.~\ref{Fig:PR_type1}. Note that $1\le  \mathrm{PR}_{E_k}\le 3$.

We saw above that the ground state energy $E_0\to -\infty$ as $y\to\infty$, while $E_0\to\eps_0$ for $y\to0$. Let $y$ be large enough that $E_0$ is well separated from $\eps_0$. Then,
we can replace summation in \erefo{ex} with integration and obtain
\beg
\ln \frac{E_0-w}{E_0+w}=\frac{\delta}{y}=\frac{2w}{Ny}.
\label{E}
\en
Performing the same replacement in \erefo{pr1} and using \erefo{E}, we derive
\beg
\mathrm{PR}_{E_0}=\frac{3 N  }{1+2\cosh(\delta/y) }.
\label{pr00}
\en
Note that in the limit $y\to\infty$, $\mathrm{PR}_{E_0}=N$ in agreement with \erefo{pr0}. This expression also allows us to estimate the  value $y_c$ beyond which the ground state becomes extended. We obtain $\lambda_c =y_c/\delta\approx 1/\ln N$. This also corresponds to the coupling for which the gap in the spectrum $\Delta=E_1-E_0\approx -w-E_0$ becomes comparable to the spacing $\delta$. For a superconductor described by the BCS model \re{bcs} this localized-extended crossover translates into a normal-superconducting 
one\cite{anderson1959,imry2003}. As $N\to\infty$ this crossover becomes a quantum phase transition at $\lambda=0$, i.e. any infinitesimal coupling is sufficient to make the ground state extended (superconducting).  The localized character of the excited states for the specific case of $\gamma_i=1$ has been demonstrated in a previous work as well~\cite{ossipov}.

\section{A model with finite-ranged hopping}             
We now consider the following Anderson-type model in one dimension with nearest neighbor hopping.
\begin{eqnarray}
H&=&\sum_{i}\epsilon_{i}n_{i} -yt\sum_{i}(c^{\dag}_{i}c_{i+1}+h.c.) \nonumber \\
 &=&H_{0}+yH_{1}.
\label{Eq:Hamiltonian}
\end{eqnarray}
This corresponds to the case with $t_{ij}=t$ for $|i-j|=1$ and $0$ otherwise for the general Hamiltonian in \erefo{Eq:genham}. $H_0$ is the zeroth order Hamiltonian with only the on-site potential and $H_1$ contains the hopping.  It is known that all single particle eigenstates of this Hamiltonian are localized~\cite{anderson.1958,tvr.1979}.
\subsection{Construction of the conserved charges}
Proceeding as for the case of Type-1 Hamiltonians, we focus on the conserved charge $Q_0$, corresponding to the site $i=0$, which to lowest order is equal to $n_0$. However, in this case $Q_0$ is not simply linear in $y$.   In fact, it can be argued that the an expansion of $Q_0$ in the hopping does not truncate at any finite order in the thermodynamic limit. Indeed, as explained in the Introduction, conserved charges  are generally infinite power series in $y$. We thus assume  $Q_i$  of the form
\begin{eqnarray}
Q_i=P_{i 0}+yP_{i 1}+y^2P_{i 2}+\dots,
\label{Eq:comhop}
\end{eqnarray}
where $P_{i 0}=n_0$ and $P_{i 1},P_{i 2} \dots$ are operators to be determined in terms of the microscopic parameters  subject to the condition  $[Q_i,H]=0$. For concreteness, we first  take our one dimensional system to be a finite-sized ring of $N+1$ sites going from $0$ to $N$.  

\begin{figure}
\begin{center}
\includegraphics[width=3in]{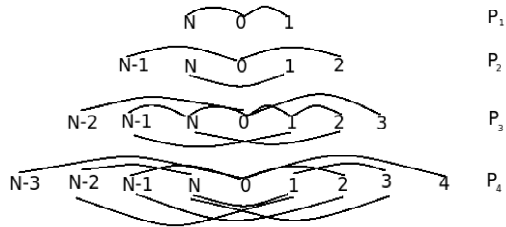}
\vspace*{-2.8in}
\end{center}
\caption{Schematic diagram  showing   hopping terms   present in the operators $P_1 - P_4$.
The base site $0$ is in the middle and its neighbors are sites $1$ and $N$, since we imposed periodic boundary conditions. 
Lines connecting pairs of sites  indicate the presence of the corresponding hopping term in the operator $P_m$.  
Note that the range of the hopping in   $P_m$ increases with $m$.}
\label{Fig:recurrance1}
\end{figure}

Since the Hamiltonian $H$ and and all the zero order charges $n_i$ are quadratic in the creation and annihilation operators, we take all the operators $P_{i1},P_{i 2} \dots$  to be  similarly quadratic, i.e.
\begin{equation}
P_{i m}=  \sum_{jk }\eta^{(m)}_{jk}(i) c_{k}^{\dag}c_{j},
\label{Eq:pform}
\end{equation}
where the  symmetric coefficients $\eta_{jk}^{(m)}(i)=\eta_{kj}^{(m)}(i) $ are to be determined.  We have
$$
[Q_i,H]=[P_{i 0},H_{0}]+\sum_{m}y^{m+1}\left([P_{i m},H_{1}]+[P_{i m+1},H_{0}]\right).
$$
The requirement that the commutator vanishes to all orders in $y$  requires 
\beg
 \begin{split}
 [P_{i m},H_{1}]+[P_{i m+1},H_{0}]=0, \end{split}
\label{com}
 \en 
and yields a recursion relation among  $\eta's$. 
 
\begin{eqnarray}
 \eta^{(m+1)}_{ab}(i)  =  \delta_{ab} R^{(m+1)}_a(i)  
 + \frac{1-\delta_{ab}}{\epsilon_a-\epsilon_b}\sum_j [(t_{aj} \eta^{(m)}_{jb}(i)- \eta^{(m)}_{aj}(i) t_{jb}], \label{rec}
\end{eqnarray}
with initial conditions  $\eta_{ab}^{(0)}(i)=\delta_{i a} \delta_{i b}$. The diagonal term $R_a^{(m+1)}$ represents  a ``gauge'' freedom,  since the corresponding term in  $P_{i m}$ commutes trivially with $H_0$. We  further discuss this freedom  below. Specializing to the case of nearest neighbor hopping Eq.~(\ref{Eq:Hamiltonian}) and with $i=0$, %
 it can be verified that terms present in $P_{0 m}$ are of the form
\begin{itemize}
 \item $\eta_{0m}^{(m)}(c^{\dag}_{0}c_{m}+c_{m}^{\dag}c_{0})$
 \item $\eta_{0,N-(m-1)}^{(m)}[c^{\dag}_{0}c_{N-(m-1)}+c_{N-(m-1)}^{\dag}c_{0}]$
\item $\sum_{ij \neq m, |i-j|=\mathrm{even}\leq m} \eta_{ij}^{(m)}(c_{i}^{\dag}c_{j}+c_{j}^{\dag}c_{i})$ (if $m$ is even)
\item $\sum_{ij \neq m, |i-j|=\mathrm{odd}\leq m} \eta_{ij}^{(m)}(c_{i}^{\dag}c_{j}+c_{j}^{\dag}c_{i})$ (if $m$ is odd)
 \end{itemize}
This is shown schematically in Fig.~\ref{Fig:recurrance1} for the first few $P_{m}$.

The $Q_i's$  are related to each other  by translating all  site indices in the above relations by an appropriate number. By construction,  they all  commute with $H$. Since
$H$ is generally non-degenerate, this implies $Q_i$ also commute among themselves, $[Q_i, Q_j]=0$ $\forall$ $i,j$. To see this, first recall that for Hermitian matrices $[A, B]=[A, C]=0$
implies $[B,C]=0$ as long as eigenvalues of $A$ are non-degenerate.  All operators involved in the above construction of $Q_i$ are of the form
$\hat A=\sum_{ij} A_{ij}c_{i}^{\dag}c_{j}$, where $A_{ij}$ is a Hermitian $N\times N$ matrix, which represents  operator $\hat A$ in the sector with total particle number $n=1$. Moreover, the commutativity of any two such operators is equivalent to that of the underlying matrices. Eigenvalues of the Hamiltonian in the $n=1$ sector at $y=0$ are $\epsilon_i$, which are assumed to be distinct, i.e. the corresponding matrix is non-degenerate at $y=0$. By continuity of the eigenvalues in $y$,  it remains non-degenerate in some finite interval (until the first level crossing) of the real axis containing $y=0$. Thus, $[Q_i, Q_j]=0$ $\forall$ $i,j$ in this interval of $y$. But, as can be seen e.g. from the above construction of $Q_0$, commutativity of $Q_i$  on any finite interval of values of $y$ implies that they commute for all $y$.

We noted above  that commutation relations \re{com} and consequently recursion relations \re{rec} do not constrain the diagonal part of the coefficients $\eta^{(m)}$, i.e. $R^{(m)}_{r}$, for $m\ge 1$. The choice of  $R^{(m)}_{a}(i)$ however {\em does} affect the off-diagonal part of $\eta^{(k)}$ for $k>m$. In our construction of $Q_i$ we set $R^{(m)}_{a}(i)=0$ for all $m\ge 1$, since this leads to the most compact description of these objects.  we will refer to this as the {\em standard gauge}.
 Conserved charges $\widetilde{Q}_i$ resulting from {\em any other choice} $R^{(m)}_{a}(i)$ uniquely relate to our standard gauge $Q_i$'s, a brief calculation shows their relationship is
\beg
\widetilde{Q}_i=Q_i+\sum_m y^m \sum_r R^{(m)}_r(i) Q_r. \label{eq23}
\en
Another  advantage of our choice of a gauge is  in a simple relationship between the Hamiltonian   (\ref{Eq:Hamiltonian}) and the conserved charges, namely,
\beg
H=\sum_i \epsilon_i Q_i.
\label{hq}
\en
To see this, consider the difference
\beg
H-\sum_i \epsilon_i Q_i=y W_1+y^2W_2+y^3W_3+\dots,
\label{inter}
\en
where $W_i$ are $y$-independent operators. Note that the zeroth order term cancels in the difference. Since $H$ commutes with all $Q_i$, the right hand side (RHS) of \erefo{inter} must also commute. This implies in particular  $[W_1, n_i]=0$ for all $i$ (from the coefficient at the lowest power of $y$ in the commutator of the RHS with $Q_i$), which in turn means 
that $W_1=\sum_i r^1_i n_i$. Now note that   the left hand side (LHS) has zero diagonal matrix elements, i.e. no terms of the form $c_r^\dagger c_r$. This is because the zeroth order term is absent, while higher order terms have no diagonal matrix elements since $\eta^{(m)}_{rr}=0$ for all $m\ge 1$ in our gauge (and similarly the diagonal is absent in other $Q_i$). Then, the diagonal matrix elements must vanish on the RHS as well, to all orders in $y$. In particular, $r^1_i =0$, i.e. $W_1=0$ and
\beg
H-\sum_i \epsilon_i Q_i=y^2W_2+y^3W_3+\dots
\label{inter1}
\en
Applying the same argument to the RHS of this equation we similarly obtain $W_2=0$ etc., until we finally arrive at \erefo{hq}.

 \subsection{Type-1 Hamiltonians redux}
  We have seen above that  the conserved charges are power series in the hopping. This is unlike the case of Type-1 Hamiltonians, where the power series {\em truncates} after the first term. The gauge where the series  truncates corresponds to having distinct terms for $m=1$, one can  see in  Eq.~(\ref{Eq:genqtypeI})  (the gauge terms are indicated in the lower braces). 

It is an amusing exercise to determine the correct gauge terms that lead to truncation, starting from the recursion relations    Eq.~(\ref{rec}). To obtain Type-1 Hamiltonians we set $t_{ij}= \gamma_i \gamma_j$, so that the recursions simplify to
\begin{eqnarray}
\eta_{ab}^{(m+1)}(i) &=&\delta_{ab} R_{a}^{(m+1)}(i)- \frac{1- \delta_{ab}}{\epsilon_a-\epsilon_b} \left( Y^{(m)}_{ab}(i)- Y^{(m)}_{ba}(i)\right) \nonumber \\
Y^{(m)}_{ab}(i)&=& \sum_j \eta^{(m)}_{a j}(i) \gamma_j \,  
\gamma_b. \label{eq27}
\end{eqnarray} 
With the initial condition $\eta_{ab}^{(0)}(i) = \delta_{i a} \delta_{i b}$, we obtain at the first level
\begin{eqnarray}
\eta_{ab}^{(1)}(i) &=&\delta_{ab} R_{a}^{(1)}(i) + \frac{\gamma_a \gamma_b (1-\delta_{ab})}{\epsilon_a-\epsilon_b} \, \left( \delta_{i b} -\delta_{i a}\right). \label{eq28}
\end{eqnarray}  
At this point we pause and ask if  we can choose  the gauge term $R_a^{(1)}(i)$ such that $\eta_{ab}^{(2)}(i)$ can be made to vanish identically, so that the iterations stop at the first level.
From Eq.~(\ref{eq27}) we see that the relevant condition is the vanishing of $ \left( Y^{(1)}_{ab}(i)- Y^{(1)}_{ba}(i)\right)$. Using Eq.~(\ref{eq27}) compute 
\begin{equation}
Y^{(1)}_{ab}(i)= \gamma_a \gamma_b \{R_{a}^{(1)}(i) + \frac{\gamma_i^2}{\epsilon_a-\epsilon_i} (1- \delta_{ia}) - \delta_{ia} \sum_j' \frac{\gamma_j^2}{\epsilon_j-\epsilon_i} \}. \label{eq29}
\end{equation} 
We may choose  $R^{(1)}$ so  that  the term in braces  vanishes, thus leading to the truncation of the iterations.   From Eq.~(\ref{eq28}) we have the complete first order term, and we can proceed to  construct the  charge (denoting the currents by the symbol $\widetilde{Q}$)
\begin{equation}
\widetilde{Q}_i   = n_i + y \sum_{ab} \eta_{ab}^{(1)}(i) c^\dagger_a c_b,
\end{equation}  
which is identical to that in Eq.~(\ref{Eq:genqtypeI}).

The use of the gauge term here is very special, and guided by our understanding of this model. On the other hand,  we could by default set all the gauge terms $R^{(m)}$ to zero, giving us the irreducible (i.e. standard gauge) currents. These no longer truncate even for Type-1 Hamiltonians.
For completeness we note the second order term for the current in this (standard) gauge 
\begin{eqnarray}
{Q}_i & = & n_i + (y + y^2 \sum_j' \frac{\gamma_j^2}{\epsilon_j -\epsilon_i})\times
 \sum_{i \neq j} \frac{\gamma_i \gamma_j}{\epsilon_j -\epsilon_i}  (c^\dagger_i c_j+c^\dagger_j c_i)
\nonumber \\
&&  + y^2   \gamma_i^2 \times \sum'_{a,b} \frac{\gamma_a \gamma_b}{(\epsilon_a-\epsilon_i)(\epsilon_a-\epsilon_j)} c^\dagger_i c_j +O(y^3). \label{eq31}
\end{eqnarray}
Thus the Type-1 Hamiltonians allow for  variety of expressions of the constants of motion.
To establish  their equivalence in general is a subtle  problem, where  some surprising results have been found quite recently in Ref.~\cite{owusu2013functionally}. 
  
 This type of gauge choice, made explicit in our construction  could  be exploited  further to test the  possibility that  the series can  take simpler forms, as compared to a  brute force expansions to infinite order. We leave  this  interesting question for future investigation. 

\subsection{Currents found from the  Rayleigh Schr\"{o}dinger (locator) expansion}
 A natural question that  arises is the relationship between  the currents found above and those found  from a brute force expansion of the projection operators of the Anderson model in powers of the coupling constant $y$. The model has a  formal single particle eigenfunction expansion  in the form
\beg
|\Psi(y)\rangle = \sum_k u_{0 k}(y) c^\dagger_k |0\rangle,
\en
with an  initial condition  localized say at the  site $0$ as $u_{0k}(0)= \delta_{k 0}$. 
The projector $ Q=|\Psi(y)\rangle \langle \Psi(y)|$ can be expanded in a series in $y$
\beg
\hat{Q} =\sum_{j,k} u_{0 j} u^*_{0 k} c^\dagger_j c_k  =  \hat{P}(0)+ y \hat{P}(1)+y^2 \hat{P}(2)+\ldots  \label{eq-28}
\en
so that the basic expansion of the wave functions in a Rayleigh Schr\"{o}dinger (RS) series in $y$ generates the conserved currents.  We can use the standard result in text books \footnote{For  e.g. see Eq.~(5.1.44)   {\em Modern Quantum Mechanics},   J J Sakurai (Pearson Education 1994))} to write a perturbative expansion for the state at site $0$ with standard normalization to $u_{00}=1$  as
\beg
|\Psi(y)\rangle = c^\dagger_0 |0\rangle +\sum_{k \neq 0} u_{0 k} c^\dagger_k |0\rangle, 
\en
with a power series expansion for $u_{0k}$
\begin{eqnarray} u_{0k} = - y \frac{t_{0k}}{\epsilon_0-\epsilon_k} + y^2 \sum_{ l \neq 0}
\frac{t_{k l} t_{l 0}}{(\epsilon_0 - \epsilon_k) (\epsilon_0 - \epsilon_l)} 
 - y^2  \frac{t_{00} t_{k 0}}{(\epsilon_0-\epsilon_k)^2}+ O(y^3). \label{eq-30}
\end{eqnarray}
Using this expansion, we may  generate the series Eq.~(\ref{eq-28}), the result is explicitly stated below in Eq.~(\ref{final}).
  From this series we can verify  to second order, that this series   differs from that in the standard gauge Eq.~(\ref{Eq:comhop}) by  specific  gauge  terms. The advantage of Eq.~(\ref{Eq:comhop}) is that this gauge invariance is manifest in the construction by the nested commutators.  On the other hand, Eq.~(\ref{eq-28}-\ref{eq-30}), corresponds to  a particular gauge picked out by the R-S method, and the currents found here are some linear combinations of the ones in  Eq.~(\ref{Eq:comhop}) as in Eq.~(\ref{eq23}).  

It seems to us that  the series in Eq.~(\ref{Eq:comhop}) possesses  an essential simplicity  relative to the Rayleigh-Schr\"{o}dinger series Eqs.~(\ref{eq-28}-\ref{eq-30}). The R-S perturbation expansion simultaneously  determines the energy eigenvalue, and  for this purpose  very specific gauge terms are needed. On the other hand all terms in  Eq.~(\ref{Eq:comhop}) are generated by completely off diagonal terms, those terms  that avoid multiple visits to any  site. This leads to simpler recursion relations, as in Eq.~(\ref{rec}), relative to the RS series. For this reason our numerical work in this paper uses the series in Eq.~(\ref{Eq:comhop}).

\subsection{Locator expansion for Type-1 Hamiltonians}

The Rayleigh-Schr\"{o}dinger series can also be constructed for Type-1 Hamiltonians using the exact 
eigenstates $|E\rangle$ with eigenvalues $E$ as given in \erefo{state} and \erefo{ex}.
The projector $|E\rangle\langle E|=\sum_{ij}\frac{\gamma_i\gamma_j c_i^\dag c_j}{(E-\eps_i)(E-\eps_j)}$ can be expanded in $y$ as shown in \erefo{eq-28}.
 In the limit $y\to 0$, the roots of \erefo{ex} tend to $\eps_i$. We take the root $E\to \eps_0$ to obtain 
$\widetilde{Q}_0$ the conserved charge corresponding to site $0$ calculated using the Rayleigh-Schr\"{o}dinger gauge. Other roots yield other $\widetilde{Q}_i$. 
Expanding \erefo{ex} for $E$  in $y$ near $E=\eps_0$, we get
\beg
E=\eps_0 -y\gamma_0^2+y^2\gamma_0^2\sum_{i\ne0}\frac{\gamma_i^2}{\eps_0-\eps_i} +O(y^3).
\label{expan}
\en
Since, the projector diverges in $y\to0$ limit, we define our conserved charge as $\widetilde{Q}_0=\frac{(E-\eps_0)^2}{\gamma_0^2}|E\rangle\langle E|$ to make it well behaved. $\widetilde{Q}_0$ is given by, 
\begin{eqnarray}
\widetilde{Q}_0=n_0+&&\frac{E-\eps_0}{\gamma_0}\sum_{j\ne0}\frac{\gamma_j(c_0^\dag c_j+c_j^\dag c_0)}{E-\eps_j} \nonumber\\
&+& \frac{(E-\eps_0)^2}{\gamma_0^2}\sum_{i,j\ne0}\frac{\gamma_i\gamma_j c_i^\dag c_j}{(E-\eps_i)(E-\eps_j)}
\label{long}
\end{eqnarray}
Then, replacing 
$\frac{(E-\eps_0)^2}{\gamma_0^2}\to y^2\gamma_0^2$ and then $E\to \eps_0$, we have obtained $\widetilde{Q}_0$ as a combination of 
$Q_i$ (see Eq.~\ref{Eq:genqtypeI})  as follows,
\beg
\widetilde{Q}_0 =Q_0 -y \sum_{i\ne0}\frac{\gamma_0^2 Q_i +\gamma_i^2 Q_0}{\eps_0-\eps_i} +O(y^{3})
\label{final}
\en
Other $\widetilde{Q}_k$ can be obtained with the replacement $0\to k$. Unlike $Q_k$, there is no indication of the series truncating at any finite order for $\widetilde{Q}_k$.

\subsection{Convergence of the power series}
\label{convergence}

 The conserved charges constructed above depend on the microscopic parameters of the Hamiltonian, i.e. the hopping and on-site energies. As we shall show later, the same Hamiltonian can have a localized and delocalized phase depending on the values of these parameters. It is thus important to understand if and how the conserved charges themselves differ in the two phases. More precisely, how do the conservation laws ``know'' whether a particular choice of microscopic parameters produces a localized or delocalized phase? 

The answer has to do with their convergence since they are expressed as power series in the microscopic parameters and particle operators. We thus need to state in what sense the power series are convergent. A reasonable condition for convergence is a sufficiently rapid decay of the coefficients $\eta_{ij}^{m}$ with increasing $m$. However, this is complicated by the fact that there are energy difference denominators in the coefficients $\eta_{ij}^m$ that can cause them to blow up when the on-site energies at two different sites are equal. To avoid this, we restrict ourselves to a  particular type of disorder that may be termed "non-resonant". By this we mean any ensemble  of $\epsilon_i$, which shows  ``level repulsion'', i.e.  the probability of finding  $\epsilon_i$ very close to each other is very small. 

From the  random matrix theory, we know that the eigenvalues of a generic matrix display level repulsion in their eigenvalues of various degree, the Gaussian Orthogonal Ensemble (GOE)~\cite{mehta.2004}
 of real symmetric matrices has the least level repulsion. This condition  ensures that perturbative resonances from small denominators,  that would otherwise cause individual terms in the expansions of the conserved charges to diverge,  are prohibited. This choice is similar to the one involving limited level attraction recently adopted in the context of many-body localization~\cite{imbrie.2014}. 
 
 We have verified that this distribution of onsite energies gives us localization (as indicated from a calculation of the participation ratio) immediately upon switching on the hopping term. Thus, this particular choice of onsite energies, which is of great convenience from the point of view of calculations, is also not unphysical.
\begin{figure}
\begin{center}
\includegraphics[width=2.5in,angle=-90]{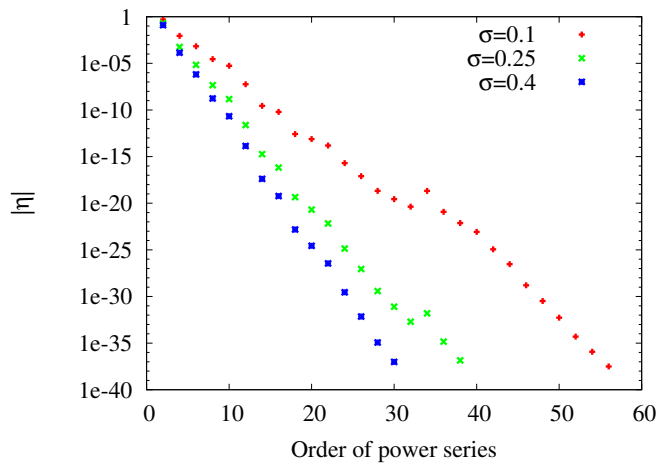}
\end{center}
\caption{Plot indicating convergence of   conserved charges [see Eqs.~(\ref{Eq:comhop}) and (\ref{Eq:pform})] of the Anderson model (\ref{Eq:Hamiltonian}) for $N=500$.  $|\eta|$ 
represents a typical $m$-th coefficient $\eta_{ij}^m$ averaged over a distribution of the on-site disorder $\epsilon_i$, see the end of Sect.~\ref{convergence}. The plot shows the logarithm of the average as a function of $m$. $\epsilon_i$ are drawn 
from the eigenvalues of real symmetric matrices whose elements are
Gaussian  random variables of variance $\sigma=$0.1, 0.25 and 0.4, and we set $yt=1$.}
\label{Fig:GOE_convergence}
\end{figure}
The on-site energies $\epsilon_i$ are  drawn from the eigenvalues of a real symmetric matrices whose elements are taken from 
a Gaussian random distribution with fixed variance. 
The eigenvalues of these matrices are assigned randomly to different sites. 
Different random assignments then constitute different realizations of disorder, which can then be averaged over
to check for convergence. The result of this procedure is shown in Fig.~\ref{Fig:GOE_convergence}, where  
$\epsilon_i$ are drawn from the eigenvalues of real symmetric matrices whose elements are taken from a Gaussian 
distribution of variance $\sigma$=0.1, 0.25 and 0.4. It can be seen that the $\eta^m$ decrease rapidly with 
increasing order of power series $m$ indicating  convergence. We have also checked the convergence of the power series for 
$\epsilon_i$ drawn from the eigenvalues of non-integrable $t-t'-V$ model, which also follow a  
GOE distribution~\cite{ranjan.2014,subroto.2006}.

Since $\eta_{ij}^{m}$ contain more than one term for each $m$, we checked the convergence of a typical term, 
which is of the form 
$\frac{t^{m}}{(\epsilon_{a_1}-\epsilon_{b_1})(\epsilon_{a_2}-\epsilon_{b_2})....(\epsilon_{a_m}-\epsilon_{b_m})}$. Recall that the $m^{\rm th}$ order term in the calculation of $\tilde{Q}_0$ involves sites with labels between $N-(m-1)$ and $m$ as can be seen from Fig.~\ref{Fig:recurrance1}. Thus, the only values of $\epsilon_i$ involved are are those chosen from $[\epsilon_{N-(m-1)},\epsilon_m]$ ($\epsilon_0$ is at the center) such that $\epsilon_{a_i} \neq \epsilon_{b_i}$, $\forall i$ and $\max|a_{i}-b_{i}| = m$

\begin{figure}
\begin{center}
\vspace*{-0.1in}
\includegraphics[width=2.4in,angle=-90]{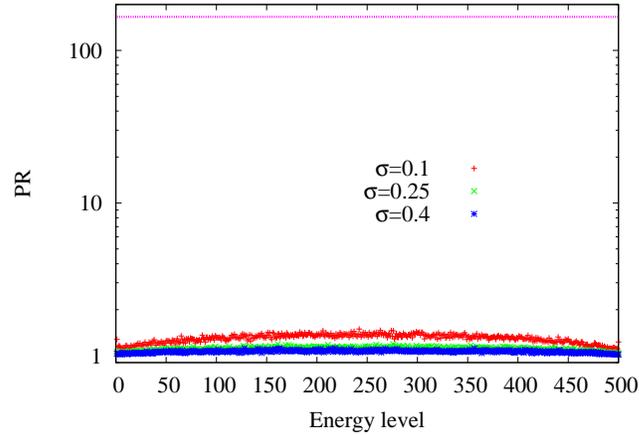}
\end{center}
\caption{ PR of eigenstates of the Anderson model (\ref{Eq:Hamiltonian}) for  $N=500$ numbered in ascending order according to the energy levels.  On-site disorder  $\epsilon_i$ is drawn from the eigenvalues of real symmetric matrices whose elements are
Gaussian  random variables of variance $\sigma=$0.1, 0.25 and 0.4, and we set $yt=1$.
  Blue dashed line corresponds to the typical value of PR in delocalized phase.}
\label{Fig:PR}
\end{figure}
As the aim of this work is to construct conserved charges in localized systems, it is legitimate to ask whether this slightly non-standard choice of disorder distribution produces localization. We have verified this through numerical exact diagonalization by calculating the 
PR.  We find that the PR for different eigenstates is indeed close to zero for systems of size $N=500$ as shown in Fig.~\ref{Fig:PR}, consistent with 
  localization. We thus conclude that our model with on-site energies taken from a GOE distribution does indeed produce a localized phase. 
\begin{figure}
\begin{center}
\vspace*{-4.5in}
\includegraphics[width=5.0in]{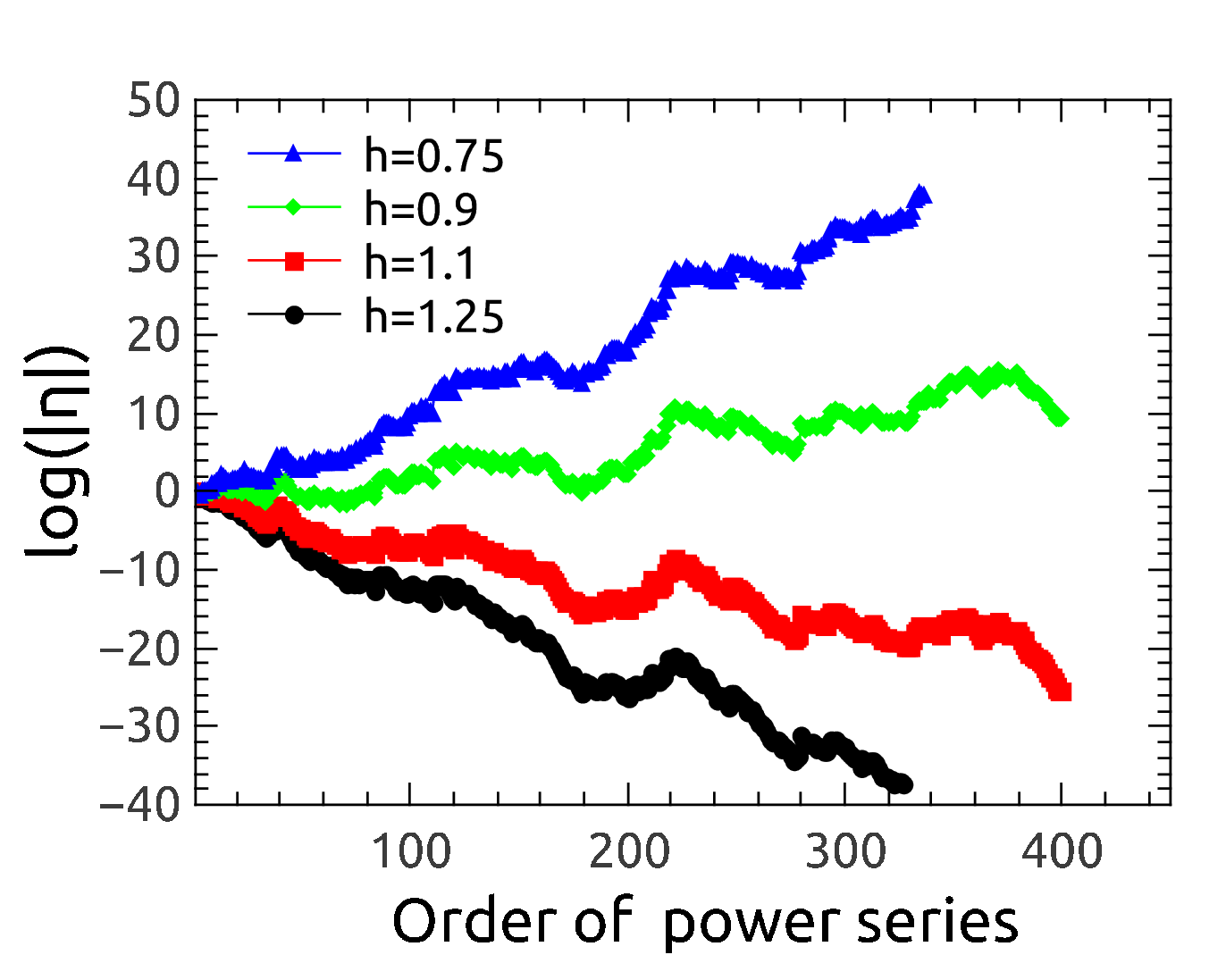}
\end{center}
\caption{ Conserved charges   for Aubrey-Andre model converge for $h>1$ (localized phase) and diverge for
$h<1$ (delocalized phase). Here $|\eta|$ represents a typical $m$-th coefficient $\eta_{ij}^m$ in \erefo{Eq:pform} [see the end of Sect.~\ref{convergence}],
$N=900$ and $\beta=\frac{\sqrt{5}-1}{2}$. The plot shows $\log|\eta|$ as a function of $m$.}
\label{Fig:convergence_AA}
\end{figure}
A similar exercise to construct the conservation laws for the above model has been 
carried out in Ref.~\cite{ros.2015}. In that work too, the conserved charges have been 
constructed as infinite operator series but whose coefficients correspond to the amplitudes of a 
particle to be on the sites of a square lattice whose sides are the physical one dimensional lattice. 
The recursion relation obtained is between conserved charges on different sites and the convergence of 
the series is assumed to follow from the exponential decay of the eigenfunctions of the Hamiltonian. 
In our calculations, we construct the conserved charges directly in terms of the microscopic parameters 
of the Hamiltonian and our convergence criterion is not based on  any assumption about the nature of the 
eigenstates of the Hamiltonian. In fact, as we show in the next section, the convergence of the series for 
the conserved charges can be used to identify the delocalized and localized phases instead of the eigenfunctions.

\section{Aubry-Andre model}

Having constructed the conserved charges for a model with finite-range hopping and defined a condition for convergence of the power series for them, we can 
further investigate the meaning of our convergence criterion. In particular, since our goal is to identify the validity of our  construction of the conservation laws with the presence of localization, the power series should fail to converge according to our criterion in a delocalized phase. 

We thus require a non-interacting model with disorder in one dimension which has a  delocalized phase. 
While any model with finite-range hopping and an on-site random potential in one dimension always produces 
localization~\cite{anderson.1958,tvr.1979}, a quasi-periodic potential can produce localized and delocalized phases. Such a model is the Aubry-Andre model~\cite{aubry.1980} given by the Hamiltonian  
\beg 
 H=h\sum_{j}\cos(2\beta\pi j)c^{\dag}_{j}c_{j} -\frac{1}{2}\sum_{j}(c^{\dag}_{j}c_{j+1}+\mathrm{h.c.}),  
 \label{Eq:Aubry}
\en 
where $\beta$ is an irrational number. The parameter $h$ can be tuned to effect a transition from a localized phase (for $h > 1$) to a delocalized phase (for $h<1$)~\cite{aubry.1980}.  We note that this model is usually studied with an additional term that introduces a
$p$-wave pairing gap~\cite{wang.2013}, but we set it equal to zero for our analysis. 

The localized phase here is one in which {\em all} single particle states are localized and similarly {\em all} single particle states are delocalized in 
the delocalized phase. The transition between these phases happens at $h=1$. 
Since the Hamiltonian in \erefo{Eq:Aubry} is also of the form  \re{Eq:Hamiltonian}, we can use the expressions obtained 
for the $\eta^{m}_{ij}$ in the previous section to construct the conserved charges. These will now 
depend on the parameter $h$ (i.e. $y\to (2h)^{-1}$ in the previous section) and if the criterion for 
convergence postulated by us is a valid one to detect localization, we should observe the power series to 
converge in the localized phase ($h>1$) and diverge in the delocalized phase ($h<1$). 
This  is indeed  the case as we see e.g. from Fig.~\ref{Fig:convergence_AA},  which shows that 
a typical matrix element of  $\eta^m$ goes to zero quite rapidly with increasing $m$ for $ h>1$ but diverges for $h<1$.
Thus, we have established that our convergence criterion is valid for identifying the   localization--delocalization transition. 

\section{Interactions}

We now turn to systems with interactions. The simplest way to introduce interactions to  models we studied here is through a nearest neighbor density-density term. Let us, for example, add such a term to \erefo{Eq:Hamiltonian},
\begin{eqnarray}
H&=&\sum_{i}\epsilon_{i}n_{i} -ty\sum_{i}(c^{\dag}_{i}c_{i+1}+h.c.)+V\sum_{i}n_{i}n_{i+1} \nonumber \\
&=&H_{0}+V\delta H,
\label{Eq:Hamint}
\end{eqnarray}
where we redefined $H_0$ as compared to  \erefo{Eq:Hamiltonian}.

We assume that the particles here are spineless fermions. It is tempting to try a construction of the conserved charges starting from a zeroth order Hamiltonian that combines the on-site and interaction terms since they commute with each other and their eigenstates are localized at every site. However, the interaction term is quartic in  creation and annihilation operators and so the conserved charges can no longer be assumed to be power series in the hopping with each term  quadratic in the creation and annihilation operators. Such an assumption leads to no solution for the coefficients since the commutators keep producing terms with increasingly longer trails of creation and annihilation operators as one goes to higher orders in the hopping. 
A more profitable exercise is to try to obtain the conserved charges as power series in the hopping but only to the first order in the interaction. While these are not exact, they offer a reasonable approximation in the limit of small interaction strength. Weak interactions typically should not 
destroy the localization present in the non-interacting limit and thus conserved charges should continue to exist. 

We know from our previous calculation that the operator of the form $Q_{0}=n_{0}+\sum_{ijm}\eta_{ij}^{m}y^{m}c_{i}^{\dag}c_{j}$
commutes with $H_0$. Let us  now define a new operator $Q=Q_{0}+V\delta Q$ to linear order in $V$ and calculate the 
commutator.
\begin{eqnarray}
[Q,H]&=&[Q_{0}+V\delta Q, H_{0} +V \delta H] \nonumber \\
&=& V\left([\delta Q,H_{0}]+[Q_{0},\delta H]\right)+ O(V^{2}).
 \end{eqnarray}
 We choose $\delta Q$ such that $[\delta Q,H_{0}]+[Q_{0},\delta H]=0$, so that $Q$ and $H$ commute to $O(V)$. 
 We assume the form
 $\delta Q =\sum_{rstv}\psi_{rstv}c_{r}^{\dag}c_{s}c_{t}^{\dag}c_{v}$. Note this is quartic in the creation and annihilation operators since the interaction term is as well. Thus,
\begin{align*}
[\delta Q,H_{0}]+[Q_{0},\delta H]=0 \nonumber \\
\sum_{kim}y^{m}(\eta_{ik}^{m}c_{i}^{\dag}c_{k}n_{k+1}-\eta_{ki}^{m}c_{k}^{\dag}c_{i}n_{k+1}+
\eta_{i,k+1}^{m}n_{k}c_{i}^{\dag}c_{k+1} \nonumber \\
-\eta_{k+1,i}^{m}n_{k}c_{k+1}^{\dag}c_{i}) 
=ty\sum_{rstv}(\psi_{rstv-1}-\psi_{rst-1v}+\psi_{rstv+1}-\nonumber\\
\psi_{rst-1v}+\psi_{rs-1tv}-
\psi_{r+1stv}+\psi_{rs+1tv}
-\psi_{r-1stv})c_{r}^{\dag}c_{s}c_{t}^{\dag}c_{v} \nonumber \\
- \sum_{rstv} \psi_{rstv}(\epsilon_{v}-\epsilon_{t}+\epsilon_{s}-
\epsilon_{r})c_{r}^{\dag}c_{s}c_{t}^{\dag}c_{v}.
\end{align*}

We now assume that $\psi_{rstv}$ can be written as a power series in $y$, 
i.e. $\psi_{rstv}=\sum_{\alpha}A^{rstv}_{\alpha}y^{\alpha}$. Equating the coefficients at different orders of  $y$, 
one can in principle  obtain $A^{rstv}_m$ in terms of the $\eta_{ij}^{m}$ for the case with $V=0$. In fact, it 
can be seen that at a given order $m$, the $A^{rstv}_m$ are linear combinations of the $\eta_{ij}^{m}$ and the
$A^{rstv}_{m-1}$. One can also impose constraints arising from the anti-commutation of the fermionic operators,
the Hermitian nature of the conservation laws and the number of non-zero components of the $\eta_{ij}^{(m)}$ to 
severely constrain the number of non-zero components of $A^{rstv}_m$. 

Let us, for example,  derive   $\delta Q$ to the first order in $y$, i.e. we  set $m=1$. We have
\begin{align*}
 \sum_{ki}(\eta_{ik}^{1}c_{i}^{\dag}c_{k}n_{k+1}-
\eta_{ki}^{1}c_{k}^{\dag}c_{i}n_{k+1}+\eta_{i,k+1}^{1}n_{k}c_{i}^{\dag}c_{k+1}- 
\eta_{k+1,i}^{1}n_{k}c_{k+1}^{\dag}c_{i}) \nonumber \\  =
\sum_{rstv}A_{1}^{rstv}(-\epsilon_{v}+\epsilon_{t}-\epsilon_{s}+\epsilon_{r})c_{r}^{\dag}c_{s}c_{t}^{\dag}c_{v}.
\end{align*}

Since, only $\eta_{i,i+1}^{1}$ and $\eta_{i,i-1}^{1}$ are non-zero, the non-zero $A_{1}^{rstv}$ are given by the following equations:
\begin{align*}
 A_{1}^{k+1,k,k+1,k+1}=A_{1}^{k,k+1,k+1,k+1}=\frac{\eta_{k+1,k}^{1}}{\epsilon_{k+1}-\epsilon_{k}} \nonumber \\
A_{1}^{k,k-1,k+1,k+1}=A_{1}^{k-1,k,k+1,k+1}=\frac{\eta_{k-1,k}^{1}}{\epsilon_{k-1}-\epsilon_{k}} \nonumber \\
A_{1}^{k,k,k+1,k}=A_{1}^{k,k,k,k+1}=\frac{\eta_{k,k+1}^{1}}{\epsilon_{k}-\epsilon_{k+1}} \nonumber \\
A_{1}^{k,k,k+2,k+1}=A_{1}^{k,k,k+1,k+2}=\frac{\eta_{k+2,k+1}^{1}}{\epsilon_{k+2}-\epsilon_{k+1}}.
\end{align*}
The corresponding expression for $\delta Q$ to order $y$ is 
\begin{align*}
\delta Q  =  yV\sum_{rstv}A_{1}^{rstv} c_{r}^{\dag}c_{s}c^{\dag}_{t}c_{v} 
 = yV\sum_{k}\left[\frac{\eta_{k+1,k}^{1}} 
{\epsilon_{k+1}-\epsilon_{k}} (c^{\dag}_{k+1}c_{k}\right. \nonumber \\
+c_{k}^{\dag}c_{k+1})n_{k+1}  
+\frac{\eta_{k-1,k}^{1}}{\epsilon_{k-1}-\epsilon_{k}}  
(c^{\dag}_{k-1}c_{k}+c_{k}^{\dag}c_{k-1})n_{k+1} \nonumber \\
+\frac{\eta_{k,k+1}^{1}}{\epsilon_{k}-\epsilon_{k+1}} 
n_{k}(c^{\dag}_{k+1}c_{k}+c_{k}^{\dag}c_{k+1})  
+\frac{\eta_{k+2,k+1}^{1}}{\epsilon_{k+2}-\epsilon_{k+1}}n_{k} \\
\left.\phantom{\frac{\eta_{k+1,k}^{1}} 
{\epsilon_{k+1}-\epsilon_{k}}} (c^{\dag}_{k+2}c_{k+1}+c_{k+1}^{\dag}c_{k+2})\right].
\end{align*}

Other approaches to construct conservation laws for interacting systems have been 
proposed including a recent one where the interacting problem is mapped onto a non-Hermitian 
problem on a lattice in operator space~\cite{ros.2015}. A convergence criterion for the resultant 
series based on the operator norm is then used to identify localized and delocalized phases.

\section{Conclusions and discussion}

Inspired by the Type-1 Hamiltonian system, we have demonstrated a scheme to obtain the conserved charges for non-interacting 
disordered models displaying localization in one dimension. One of our motivation was an observation of  similarities between localized  and integrable systems, such as the absence of 
  level repulsion and the absence of  thermalization.  Our  conserved charges  are exhibited  as a power series in the hopping, and using a suitable  convergence criterion, we show that  the convergence (or divergence) of conserved charges tracks 
the presence (or absence) of localization. An interesting issue of ``gauge dependence'' of the conserved charges is unearthed and explored. It is shown that a full understanding of the gauge dependence  leads to considerable simplifications of the charges in some cases. On the other hand,  straightforward Rayleigh Schr\"{o}dinger perturbation theory or equivalent schemes, commit one to a particular gauge that is often inconvenient.

This work provides a novel  link between the concepts of localization and integrability. Our results hold  within the  context of the 1-d Anderson model, where all states are localized,  and the Andre-Aubry model, where (all)  states undergo a transition tuned by a coupling constant. It is not immediately obvious how to extend these results to a  higher dimensional Anderson model with a mobility edge separating the two classes of states. The Aubry-Andre model exhibits an interesting kind of duality which allows the localized and delocalized phases to be mapped onto each other with the roles of the hopping and onsite potential exchanged. The duality transformation is expressed in terms of new fermonic operators given by $c_{\bar{k}}=\frac{1}{\sqrt{L}}\sum_{n}{\exp(i2\pi\bar{k}\beta)c_{n}}$,
which are eigenstates of the momentum operator with eigenvalue: $k=\bar{k}F_{n-1}\mod F_{n}$,
where $F_{n}$ is the $n$-th Fibonacci number and $L=F_{n}$~\cite{uma.2014,huse.2013}. In terms of these fermionic 
operators the Hamiltonian \re{Eq:Aubry} becomes
\beg 
  \frac{H}{h}=  \frac{1}{h}\sum_{\bar{k}}\cos(2\beta\pi \bar{k})n_{\bar{k}}-\frac{1}{2}\sum_{\bar{k}}
  (c^{\dag}_{\bar{k}}c_{\bar{k}+1}+\mathrm{h.c.}). 
 \en

The Hamiltonian satisfies the duality relation: $H(h)/h=H(1/h)$.
We have shown that for the Aubrey-Andre Hamiltonian written in real space, one can construct set of conserved 
charges that  converge for $0<h<1$.
Because of the duality of the model one can construct similar conserved charges in terms of 
$c_{\bar{k}}$ and $c_{\bar{k}}^{\dag}$. 
The power series of these charges converge when $0< 1/h<1$ and both sets of charges diverge at $h=1$. Thus, the duality of the model allows us 
to explicitly construct conservation charges in one phase given that they exist in the other.

This can be  better understood by noting that localization is a basis dependent concept. We have been using localization (as is the standard practice) to mean 
localization in real space. To obtain the conserved charges for such a localized phase, we start from a Hamiltonian whose eigenstates are perfectly localized in real space and then add terms perturbatively in the hopping. Similarly, the delocalized phase of the Aubry-Andre model is localized in momentum space and one can then obtain its conserved charges by starting with a Hamiltonian perfectly localized in momentum space (tight binding model) and then add terms perturbatively in the on-site potential. This is the essence of the duality outlined above. Thus, the conserved charges also carry labels indicating the space (real or momentum) where the system is localized. What is important though is that once the basis in which the system is localized is identified and the conserved charges are constructed accordingly, they are sensitive to the onset of delocalization in that basis and can be used to locate localization-delocalization transitions. 

The importance of the basis can be further understood when one compares 
the behavior hard-core bosons with that of spinless fermions in the Aubry-Andre model~\cite{gramsch.2012,he.2013}. 
The duality between the localized and delocalized phases is destroyed for hard-core bosons.
As a result, the relaxation of real space local observables in the localized phase is different 
from  their conjugates in momentum space in the delocalized phase. This feature is absent for 
spinless fermions where the duality holds and as a consequence, conserved charges of the type 
derived in this work exist in both phases.

 While it is only possible to construct these charges  to lowest order in the interaction using our procedure, their fate upon the introduction of interactions can in principle be investigated numerically, which we defer to a future work.\footnote{Note that the introduction of interactions destroys the duality of the model since it  no longer has the same form in real and momentum space under the duality transform.}

\section{Acknowledgements}
The work at  UCSC was supported by the U.S. Department of Energy (DOE), Office of Science, Basic Energy Sciences (BES) under Award $\#$ FG02-06ER46319.. E. A. Y. was financially supported in part  by the David  and Lucile Packard Foundation. RM acknowledges support from the UGC-BSR Fellowship.  SM thanks the DST, Government of India and the India-Israel joint research programme for funding.

\section*{References}
\bibliographystyle{iop}

\providecommand{\newblock}{}

\end{document}